\documentclass[twocolumn,showpacs,preprintnumbers,amsmath,amssymb,aps]{revtex4-2}

\usepackage{epsfig}
\usepackage{dcolumn}
\usepackage{bm}
\usepackage[dvipsnames]{xcolor}
\usepackage[normalem]{ulem}
\usepackage{natbib,hyperref}

\usepackage{siunitx}

\usepackage{multirow}
\usepackage{adjustbox}

\usepackage{verbatim}

\begin{document}

\preprint{\today} 


\title{Measurement of the static Stark Shift of the \texorpdfstring{$7s \ ^2S_{1/2}$}{7s 2s1/2} level in atomic cesium}

\author{Jonah A. Quirk$^{1,2}$, Aidan Jacobsen$^{4}$, Amy Damitz$^{1,2}$, Carol E. Tanner$^3$ and D. S. Elliott$^{1,2,4}$}

\affiliation{%
   $^1$Department of Physics and Astronomy, Purdue University, West Lafayette, Indiana 47907, USA\\
   $^2$Purdue Quantum Science and Engineering Institute, Purdue University, West Lafayette, Indiana 47907, USA\\
   $^3$Department of Physics and Astronomy, University of Notre Dame, Notre Dame, Indiana 46556, USA\\
    $^4$The Elmore Family School of Electrical and Computer Engineering, Purdue University, West Lafayette, Indiana 47907, USA
   }

\date{\today}

\begin{abstract}

We report a new precision measurement of the dc Stark shift of the $6s\hspace{1mm} ^2S_{1/2} \rightarrow 7s\hspace{1mm}^2S_{1/2}$ transition in atomic cesium-133. Our result is 0.72246 (29) $\textrm{Hz}(\textrm{V}/\textrm{cm})^{-2}$. This result differs from a previous measurement of the Stark shift by $\sim$0.5\%. We use this value to recalculate the magnitude of the reduced dipole matrix elements $\langle7s ||r||7p_{j}\rangle$, as well as the vector transition polarizability for the $6s \rightarrow 7s$ transition, $\tilde{\beta} = 27.043 \: (36) \ a_0^3$. This determination helps resolve a critical discrepancy between two techniques for determining the vector polarizability.
\end{abstract}
\maketitle 


Measurements of extremely weak transitions mediated by the weak force interaction facilitate determinations of the weak charge predicted by the standard model~\cite{bouchiat1974parity,BouchiatB75}. The most precise measurement of the parity non-conserving (PNC) weak interaction in any atomic species to date is that of Wood \emph{et al.} \cite{WoodBCMRTW97}. This measurement yielded a ratio of the strength of the PNC interaction relative to the Stark vector polarizability, $\tilde{\beta}$.  Contention among two techniques that determine $\tilde{\beta}$~\cite{DzubaF00,BennettW1999,SafronovaJD99,VasilyevSSB02,ChoWBRW1997,TohDTJE19,TanXD2023} garner doubt about experiment and theory alike. Since theory and experiment are both critical in the determination of the weak charge of the nucleus, this discrepancy must be resolved. Recent high precision calculations of reduced electric dipole (E1) matrix elements~\cite{RobertsFG23,FairhallRG2023,TanD2023} point to two possible sources of the discrepancy: these are, the experimentally measured E1 matrix elements coupling the $7s \; ^2S_{1/2}$ state with the $7p \; ^2P_j$ states, where $j=1/2$ and $3/2$, and the theoretical values of E1 moments for $8 \le n \le 12$.  The matrix elements $\langle7s ||r||7p_{j}\rangle$ were derived from a measurement~\cite{BennettRW99} of the static Stark shift of the $7s \; ^2S_{1/2}$ state. In this paper, we report a new, high precision measurement of the Stark shift of the $7s \; ^2S_{1/2}$ state, reporting a value that differs significantly from the previous measurement~\cite{BennettRW99}.  We derive values of the E1 matrix elements $\langle 7s||r||7p_j \rangle$, which we use to show partial resolution of the $\tilde{\beta}$ discrepancy.

Upon application of a static electric field, $E$, the energy shift of an atom in an external electric field is
\begin{equation}
    \Delta U = -\frac{1}{2}\alpha E^{2},
\end{equation}
where $\alpha$ is the static polarizability of the state. We determine $\alpha_{7s}$ of the $7s$ state of atomic cesium by measuring the relative frequency shift of the $6s\hspace{1mm} ^2S_{1/2}$ and the $7s\hspace{1mm} ^2S_{1/2}$ states,

\begin{equation}
    \Delta \nu = \frac{\alpha_{6_S}-\alpha_{7_S}}{4\pi} E^{2}.
\end{equation}
(Here and through the rest of the paper $\alpha_{6s}$ and $\alpha_{7s}$ are expressed as frequency shifts per electric field squared and a factor of the reduced Planck constant, $\hbar$, is suppressed, as is common practice.  See, for example, Ref.~\cite{BennettRW99}.) When combined with precise measurements of the static polarizability of the ground state $\alpha_{6s}$~\cite{AminiG03,GregoireHHTC15}, $\Delta \nu$ can be evaluated to determine $\alpha_{7_S}$.
%
%
%
%
This frequency shift varies linearly with the applied electric field squared and can be reported as a slope,
\begin{equation}
    k=\frac{\Delta \nu}{E^2} = \frac{\alpha_{6_S}-\alpha_{7_S}}{4\pi}.
    \label{eq:k}
\end{equation}

In this measurement, we apply a large adjustable dc electric field to the atoms, and drive a Doppler-free two-photon $6s\hspace{1mm} ^2S_{1/2}\rightarrow 7s\hspace{1mm}^2S_{1/2}$ transition using the output of a 1079 nm external cavity diode laser (ECDL). This transition has a symmetric, near-lifetime-limited lineshape that does not vary with the applied dc electric field. This results in a simple accurate line center determination that is critical to measuring high precision Stark polarizabilities.



 \begin{figure*}[t!]
\begin{centering}
\includegraphics[width=0.9\textwidth]{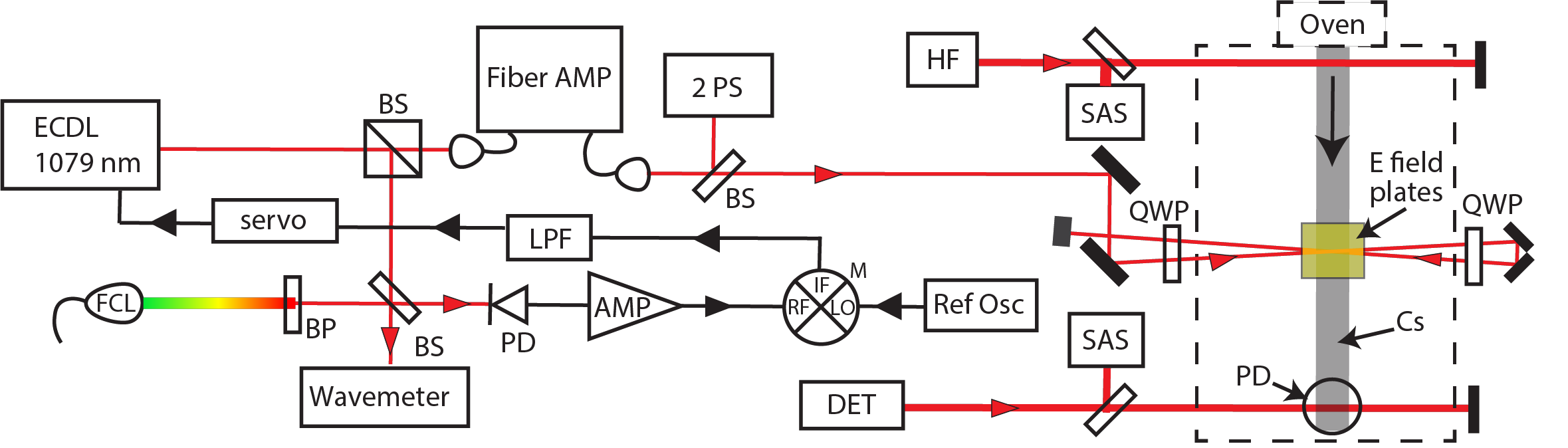}
   \caption{Experimental configuration for the $7s$ Stark shift measurement. 
   We stabilize the frequency of the 1079 nm laser (ECDL) light using an optical phase-lock loop and sweep the offset by varying the reference oscillator driving the local oscillator (LO) port of the mixer (M).   The following elements are labeled as;  BP - band pass filter, PD - photodetector, BS - beam splitter, LPF - low pass filter, SAS - saturated absorption spectroscopy cell, 2PS - two photon spectroscopy in a vapor cell, QWP -  quarter wave plate, FCL - frequency comb laser, DET - detection laser, HF - hyperfine laser. The dashed section illustrates the vacuum chamber, which contains the field plates and detection system.}
 	  \label{fig:sweepsetup}
\end{centering} 	  
\end{figure*}

\textit{Methodology} An overview of the experimental configuration is illustrated in Fig. \ref{fig:sweepsetup}. We perform measurements using an atomic beam of cesium that is generated by an effusive oven with a 1 mm diameter hole in a heated nozzle.
The atomic beam is further collimated using a 1 mm diameter aperture \num{30} cm after the nozzle. The beam travels along the length of the vacuum system, where in succession, we prepare the atoms in a single hyperfine state, drive the two-photon transition, and detect the atoms that have undergone a transition.
The cesium atoms are initially optically pumped into the $6s\: F=3$ or $6s \:F=4$ ground state using an 852 nm hyperfine laser (HF). We observe no population redistribution among the magnetic sublevels due to the hyperfine pumping.  (See \cite{supp_material} for further discussion.)
A portion of the atoms excited to the 7s state in the interaction region decays down to the emptied hyperfine level. In the detection region, these atoms are detected by driving a $6s \rightarrow 6p_{3/2}$ cycling transition from the emptied hyperfine level of the ground state, where they scatter many photons from the detection laser beam (DET). We collect these photons on a large area photo-detector, and amplify this photocurrent with a transimpedance amplifier of gain 20 M$\Omega$. The output of this amplifier is digitized and recorded.


In the interaction region, we drive the $6s \rightarrow 7s$ transition with the 1079 nm laser light generated by a commercial ECDL and fiber amplifier, with a power of 10~W.  
To reduce the Doppler broadening, we use two nearly counter propagating laser beams to excite the transition.  The crossing angle of the two unfocused beams is $\sim$10~mrad, such that the two beams overlap on the atomic beam but are separate at the edge of the chamber for beam blocking.  We reduce the two-photon absorption rate from a single propagation direction using opposite spins for the counter propagating laser beams \cite{demtroder1993laser}. 

The observed transition is primarily broadened by the lifetime (\SI{3.3}{\MHz}) \cite{TohJGQSCWE18} and residual Doppler broadening due to the small crossing angle, resulting in an observed linewidth of \num{\sim 3.8} MHz. (See \cite{supp_material} for analysis of the Doppler broadening.)  Transit time broadening is estimated to be \num{250} kHz \cite{demtroder1993laser}. Collisional effects are negligible due to operation in high vacuum, $\num{5e-7}$ Torr.

To dc Stark shift the $6s\rightarrow 7s$ transition, the cw laser light at 1079 nm intersects the atomic beam centered on a set of parallel plates used to generate an electric field. The plates are constructed of 2'' unprotected gold coated square mirrors. The mirrors are spaced using precise ceramic spacers whose length is known to within 1 $\mu$m and whose coefficient of thermal expansion is \num{7} ppm$/^{\circ}$C. Three spacers separate the gold coated mirrors and give a field plate spacing of 8.169 (1) mm, where the quantity in parentheses represents one standard error in uncertainty. We apply a potential difference of up to 5 kV to the field plates. The voltage applied to the plates is continuously monitored during data collection and deviates by less than 8 ppm during 10 scans across the two-photon transition.


The frequency of the 1079 nm laser is stabilized in an optical-phase-lock loop to a tooth of a commercial (Menlo) frequency comb laser source (FCL) whose repetition rate and carrier-envelope offset frequency are both stabilized to a GPS conditioned oscillator. The comb light is beat against the 1079 nm ECDL light. The beat signal is mixed down with a reference oscillator and low pass filtered. This signal then acts as a phase reference between the frequency comb tooth and the cw 1079 nm ECDL output. The frequency of the 1079 nm ECDL is then stabilized to offset phase lock the ECDL to the comb tooth.  Once phase locked, the ECDL attains the inherent stability and linewidth (\num{65} kHz) of the comb tooth. 

Initially, we minimize the magnetic field in the interaction region using three pairs of magnetic field coils external to the vacuum chamber while observing the Raman transition between hyperfine levels of the ground state.  We reduce Zeeman shifts among the magnetic sublevels by observing the Raman transition linewidth and effectively reducing the magnetic field to below 3 mG.

\begin{figure}
    \centering
    \includegraphics[width=0.8\linewidth]{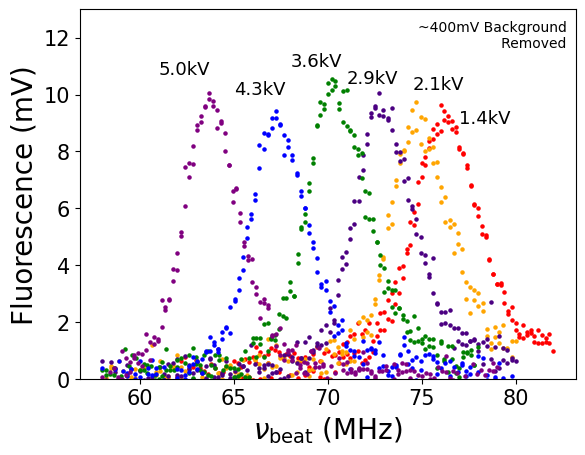}
    \caption{Single spectral scans of the $6s$ $F=3\rightarrow 7s$ $F=3$ transition for each electric field value applied. The frequency $\nu_{\rm beat}$  is the frequency difference between the ECDL and the frequency comb tooth.}  
    \label{fig:spectra}
\end{figure}

We collect $6s \rightarrow 7s$ spectra by stepping the reference oscillator frequency to vary the offset of the 1079 nm ECDL relative to the comb. At each step, we wait 50 ms to allow the signal to stabilize, and then collect 240 voltage measurements of the fluorescence on the large area photodiode at a sample rate of 480 Hz. By waiting 50 ms, greater than 25 time constants for our detection system, we eliminate any possible frequency shifts due todetector response while scanning. The fiber amplifier power is recorded as well as the beat frequency between the laser and the frequency comb source. Example spectra for a single scan at each electric field applied are shown in Fig.~\ref{fig:spectra}. Since this transition is a 2-photon transition, the actual shift in the transition frequency is twice the measured beat frequency, $\nu_{\rm beat}$ difference, and the signal strength is relatively independent of the applied static electric field. The reference oscillator is stepped 15 MHz up and down in 0.25 MHz steps in one minute. We collect ten scans at each voltage, after which the reference oscillator is advanced to the next frequency. The high voltage is alternated from a high set point to a low set point with each successive point being closer to the center voltage, 5 kV, 1.4 kV, 4.3 kV, 2.1 kV, 3.6 kV, 2.9 kV. Scans of the high voltage are also taken in the reverse order. The primary source of noise is due to Johnson-Nyquist noise from the large transimpedance gain of the photodiode amplifier. (Additional less-significant sources of noise are discussed in \cite{supp_material}.)

\begin{table}
    \centering
    \begin{tabular}{|c|c|} \hline 
           Source of error&relative size (ppm)\\ \hline \hline
           Divider ratio&50\\ \hline 
           Divider temperature& 10\\\hline
 Divider nonlinearity&5/kV\\\hline
           Spacer length&122\\ \hline 
           Plate flatness&10\\ \hline 
           Voltage measurement&22\\ \hline 
           Total&137\\\hline
    \end{tabular}
    \caption{Contribution to the uncertainty in the applied electric field.  This uncertainty (times two) constitutes the systematic error of the differential polarizability.}
    \label{tab:field_error}
\end{table}

\begin{figure}
    \centering
    \includegraphics[width=1\linewidth]{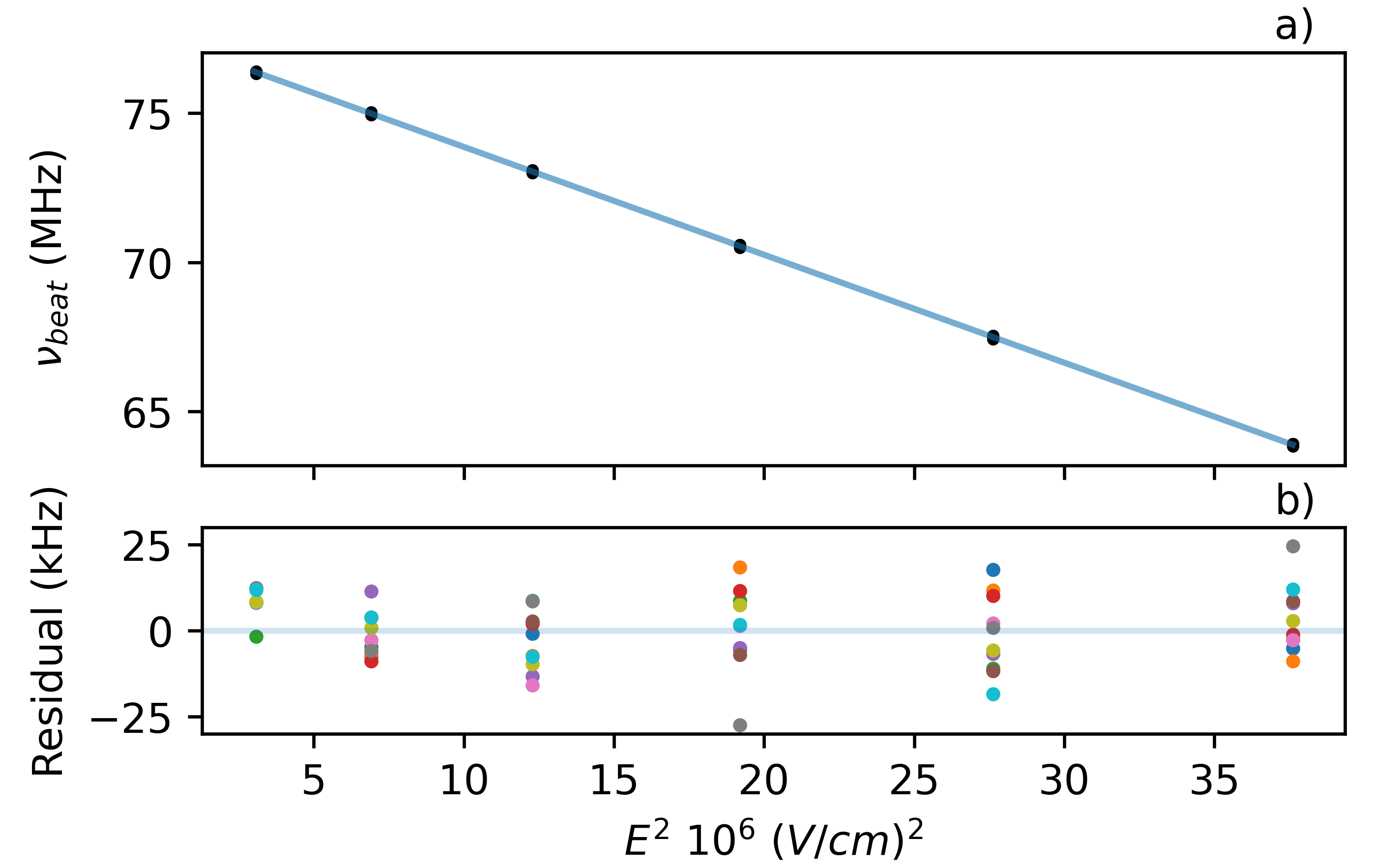}
    \caption{Fitted peak frequency relative to nearest comb tooth vs. applied electric field squared. Plot a) shows each of the 10 fitted centers averaged for each electric field for all of the runs on the $6s$ $F=3\rightarrow 7s$ $F=3$ transition. Plot b) shows the residuals of those linear fits.}
    \label{fig:fitted_centers}
\end{figure}

\textit{Analysis} We fit the spectra using a Voigt profile with the offset, amplitude, center frequency, Gaussian width, and Lorentzian width being free fitting parameters. The ten spectra are fit to determine the center frequency, and the standard error of the fitted centers is used to determine the center frequency uncertainty.  The amplitude of the $6s\hspace{1mm}F=3 \rightarrow 7s\hspace{1mm}F=3$ transition peaks were three to four times larger than those for the $6s\hspace{1mm}F=4 \rightarrow 7s\hspace{1mm}F=4$  transition.  For the ten fitted spectra on the $6s\hspace{1mm}F=3 \rightarrow 7s\hspace{1mm}F=3$  transition, the average uncertainty in the fitted linecenter is 10 kHz, while for the smaller $6s\hspace{1mm}F=4 \rightarrow 7s\hspace{1mm}F=4$, it is 15 kHz.  The fitted linewidths do not vary with the applied electric field. Spectra are collected at several applied electric fields and the field direction is reversed for several scans at each field intensity.  No effect was observed with a reversed field. The electric field squared versus center frequency is then fit with a straight line to determine the slope, the difference between the two polarizabilities $\alpha_{7s}$ and $\alpha_{6s}$. Ten slopes for the $6s\hspace{1mm}F=3 \rightarrow 7s\hspace{1mm}F=3$ transition were collected and are plotted in Fig. \ref{fig:fitted_centers}. The weighted average of these slopes is $k_{F=3}= \:$0.72267(23)$^{stat}$(20)$^{sys}$$\textrm{Hz}(\textrm{V}/\textrm{cm})^{-2}$ with a reduced chi square value of 1.68, where stat represents the statistical uncertainty (which has been expanded by the square root of the reduced chi square), and sys represents the systematic uncertainty \cite{bevington2003}, as analyzed in Table \ref{tab:field_error}. (See \cite{supp_material} for additional discussion of the determination of systematic uncertainties.) This process was repeated for the $6s\hspace{1mm}F=4 \rightarrow 7s\hspace{1mm}F=4$  transition where 15 slopes were collected whose weighted average is $k_{F=4} =\: $0.72229(32)$^{stat}$(20)$^{sys}$$\textrm{Hz}(\textrm{V}/\textrm{cm})^{-2}$ with a reduced chi square value of 1.25. We compute the weighted average in each case using $\sigma^{-2}$, where $\sigma$ is the uncertainty for each measurement, as the weight. The average Stark shift slope is $k = (7k_{F=3} + 9 k_{F=4})/16$, where 7 and 9 are the degeneracies of the $F=3$ and $F=4$ states. We also combine the systematic and statistical uncertainties in quadrature and attain a final value of $k = $ 0.72246 (29) $\textrm{Hz}(\textrm{V}/\textrm{cm})^{-2}$. The relative uncertainty of this slope is 0.04\%. We present our result, along with those of several previous experimental and theoretical studies, in Table \ref{tab:my_label}. $k$ is as defined in Eq.~(\ref{eq:k}), while $\alpha_{7s}$ is the polarizability of the $7s$ state, in atomic units.  We use here a weighted average of static polarizability measurements of the ground state $\alpha_{6s} = 401.1 \: (5) \:  a_0^3 = 0.09980 \: (11)$ Hz(V/cm)$^{-2}$~\cite{GregoireHHTC15, AminiG03} to convert between the polarizability difference between the $6s$ and $7s$ states and the static polarizability of the $7s$ state. Our value of $k$ is $\sim$0.5\% smaller than that of Ref.~\cite{BennettRW99}, with an uncertainty more than two times smaller.

%

\begin{table}[h]
    \centering
    \renewcommand{\arraystretch}{1.4}
    \begin{tabular}{|l|c|c|} \hline 
        
          &k ($\textrm{Hz}(\textrm{V}/\textrm{cm})^{-2}$)&  $\alpha_{7s}$ $(a_0^3)$\\   \hline \hline 
 This work& 0.72246(29)&\textbf{6207.9(2.4)}\\\hline
          Bennett \emph{et al.} \cite{BennettRW99} &0.7262(8)&  \textbf{6238(6)}\\ \hline 
          Watts \emph{et al.} \cite{watts1983}
&0.7103(24)&  6111(21)\\ \hline 
          Hoffnagle \emph{et al.} \cite{HOFFNAGLE1981}
&\textbf{0.7803(480)}& 6673 (386)\\ \hline \hline
          Van Wijngaarden \emph{et al.} \cite{VANWIJNGAARDEN1994}&\textbf{0.7140}&  6140\\ \hline 
          Zhou \emph{et al.} \cite{norcross1989}
&\textbf{0.7042}&  6061\\ \hline 
  Blundell \emph{et al.}~\cite{johnson1992}&\textbf{0.72572}& 6234.1\\ \hline 
  Bouchiat \emph{et al.}~\cite{BOUCHIAT198368} &0.7225& \textbf{6208}\\ \hline
    \end{tabular}
    \caption{Determinations of the static polarizability of the $7s$ state.  Calculated values are bold. Experimental determinations are above the double line and theoretical are below. See \cite{Mitroy_2010} on converting units of polarizabilities.}
    \label{tab:my_label}
\end{table}

\textit{Results} We determine the E1 reduced matrix elements $\langle 7s || r || 7p_j \rangle$ from the Stark polarizability $\alpha_{7s}$ using the sum over states expression~\cite{AngelS1968hyperfine}
\begin{equation}
    \alpha_{7s} = \frac{1}{3} \sum_n \left[ \frac{ |\langle 7s || r || np_{1/2} \rangle|^2}{E_{np_{1/2}} - E_{7s}} + \frac{ |\langle 7s || r || np_{3/2} \rangle|^2}{E_{np_{3/2}} - E_{7s}} \right].
\end{equation}
The contribution of the $7p_{1/2}$ and $7p_{3/2}$ states is by far the major term in this expression. We use the experimental matrix elements $\langle 7s || r || 6p_{j}\rangle$ determined from the $7s$ lifetime and branching ratio~\cite{TohJGQSCWE18,TohDGQSCSE19}, and theoretical values from Ref.~\cite{TanD2023} for the higher $np_j$ states, $8 \le n \le 12$.  We use state energies from the NIST data base~\cite{kramida2022nist}.  Finally, we use the theoretical value for the ratio of moments $\langle 7s || r || 7p_{3/2} \rangle / \langle 7s || r || 7p_{1/2} \rangle = 1.3891$, which is consistent across many theoretical determinations~\cite{SafronovaSC16,TanD2023,FairhallRG2023,RobertsFG23}.  The results are $ \langle 7s || r || 7p_{1/2} \rangle = 10.303 \: (3) \ a_0$, where $a_0$ is the Bohr radius, and $\langle 7s || r || 7p_{3/2} \rangle = 14.311 \: (3) \ a_0$.  The relative uncertainties of these matrix elements are 0.02-0.03\%. (Additional details of this analysis are presented in \cite{supp_material}.) These values are in very good agreement with recent theoretical determinations of these moments, as listed in Table~\ref{table:red_mat_el}.

\begin{table}[t]
    \centering
    \renewcommand{\arraystretch}{1.4}
    \begin{tabular}{|l|c|c|} \hline 
        
          & $\langle 7s || r || 7p_{1/2} \rangle$ $(a_0^3)$&  $\langle 7s || r || 7p_{3/2} \rangle$  $(a_0^3)$\\   \hline \hline 
 This work & 10.303 (3)& 14.311 (3) \\ \hline
   *Bennett \emph{et al.}~\cite{BennettRW99}&  10.325 (5)  &  14.344 (7) \\ \hline \hline      Tan \emph{et al.}~\cite{TanD2023} &10.292 (6) &  14.297 (10) \\ \hline 
    Roberts \emph{et al.}~\cite{RobertsFG23,FairhallRG2023}
&10.297 (23) &  14.303 (33)\\ \hline
  Safronova \emph{et al.}~\cite{SafronovaSC16} &10.310 (40) &  14.323 (61) \\ \hline
 Dzuba \emph{et al.}~\cite{DzubaFS97} & 10.285 (31) & 14.286 (43) \\ \hline
   \end{tabular}
    \caption{Comparison of matrix elements $\langle 7s || r || 7p_{1/2} \rangle$ and $\langle 7s || r || 7p_{3/2} \rangle$. Experimental determinations are above the double line and theoretical are below. *These matrix elements were derived from the measurements of Bennett {\emph et al.}~\cite{BennettRW99} and reported in Ref.~\cite{TohDTJE19}.}
    \label{table:red_mat_el}
\end{table}

Two prevailing techniques are used to determine the vector transition polarizability $\tilde{\beta}$ of the $6s \rightarrow 7s$ transition. The first method uses a theoretical value of the hyperfine changing magnetic dipole amplitude M1$_{hf}$~\cite{DzubaF00} and a measured value of M1$_{hf}/\tilde{\beta}$~\cite{BennettW1999} to find $\tilde{\beta} = 26.957 \: (51) \ a_0^3$~\cite{DzubaF00}. In the second technique, a sum-over-states method is used to find the scalar transition polarizability, $\tilde{\alpha}$~\cite{BlundellSJ92,VasilyevSSB02,TohDTJE19}, combined with a measured value of the ratio $\tilde{\alpha}/\tilde{\beta}$~\cite{ChoWBRW1997}.  In a recent application of this second technique~\cite{TohDTJE19}, the result was $\tilde{\beta} = 27.139 \: (42) \ a_0^3$, which showed substantial disagreement (0.67\%) with the value determined through M1$_{hf}$. Using the new results of this Stark shift measurement, and updating the theoretical values of E1 moments for $n = 8 - 12$ ~\cite{TanD2023}, we calculate $\tilde{\beta} =  27.043 \: (36) \: a_0^3$.  The difference in the values of vector polarizability $\tilde{\beta}$ determined by these two methods has been reduced to 0.29\%, less than half the previous difference.  The reduction in the value of $\tilde{\beta}$ comes from (1) the new Stark shift measurement ($\Delta \tilde{\beta} = -0.031 \: a_0^3$), (2) the improved theoretical values for E1 matrix elements for $8 \le n \le 12$~\cite{TanD2023} ($\Delta \tilde{\beta} = -0.048 \: a_0^3$), and the improved value of the valence-core and tail ($n > 12$) contributions to the polarizability~\cite{TanXD2023} ($\Delta \tilde{\beta} = -0.018 \: a_0^3$). The weighted average of this work and the value from~\cite{DzubaF00} 
for $\tilde{\beta}$, using $\sigma^{-2}$ as the relative weight factor for each, is 
\begin{equation}\label{eq:betatildeavg}
  \tilde{\beta}= 27.014 \: (30) \ a_0^3.
\end{equation}

Recent theoretical determinations of E1 matrix elements \cite{TanD2023} have been used exclusively to recalculate $\tilde{\alpha}$, which combined with the measured value of $\tilde{\alpha}/\tilde{\beta}$~\cite{ChoWBRW1997}, gives $\tilde{\beta} = 26.887\:(38) \ a_0^3$~\cite{TanXD2023}. Additional investigations are needed to bring all of these values into better agreement, but the recent theoretical results~\cite{TanD2023, TanXD2023} and the present Stark shift measurement represent a substantial improvement of the $\tilde{\beta}$ discrepancy. 


\textit{Conclusion}  In this report, we have described our precision measurement of the Stark shift of the $6s \rightarrow 7s$ transition of atomic cesium.  The precision of this measurement is facilitated by locking the laser frequency to a tooth of a stable frequency comb laser, by the reduction of ac Stark shifts (and the associated lineshape distortion) in the interaction region, and through the use of Doppler-free two-photon absorption, which leads to narrow, symmetric spectral lineshapes.  We have analyzed our results to determine the reduced matrix elements $\langle 7s || r || 7p_{j}\rangle $, with improved precision and accuracy.  Finally, we have used this new determination of  the matrix elements to re-evaluate the scalar polarizability, $\tilde{\alpha}$ for the $6s \rightarrow 7s$ transition, as well as the vector polarizability, $\tilde{\beta}$.  The disagreement between values of $\tilde{\beta}$ determined by the two primary techniques is significantly reduced. This improved agreement in $\tilde{\beta}$ represents a critical step forward in atomic parity violation measurements.

\textit{Acknowledgment} We are grateful to A. Derevianko for helpful discussions, and for the advanced notice of his recent calculations of E1 matrix elements in cesium, with special notice to the $7s - 7p_j$ terms. 
This material is based upon work supported by the National Science Foundation under grant number PHY-1912519.  


\bibliography{biblio}

\end{document}